\documentclass[useAMS,usenatbib]{mn2e}
\usepackage{multicol}
\usepackage{psfig, epsf, epsfig}

\title[Strongly suppressed star formation]{Strong 
suppression of star formation  
and spiral arm formation in disk galaxies with counter-rotating gas disks}
\author[O. Osman  and K. Bekki,]
{Omima Osman ${}^1$\thanks{E-mail:
omima.osman@uwa.edu.au} and
Kenji Bekki${}^2$ \\ 
${}^1$
University of Khartoum - Department of Physics.
Al-Gamaa Ave, Khartoum 11115, Sudan \\
${}^2$ICRAR M468
The University of Western Australia
35 Stirling Hwy, Crawley
Western Australia 6009, Australia \\
\\}

\begin{document}

\date{Accepted, Received 2005 February 20; in original form }

\pagerange{\pageref{firstpage}--\pageref{lastpage}} \pubyear{2005}

\maketitle

\label{firstpage}

\begin{abstract}
Galaxy-wide star formation can be quenched by a number of physical processes
such as environmental effects (e.g., ram pressure stripping) and supernova feedback.
Using numerical simulations,
we here demonstrate that star formation can be severely suppressed in disk galaxies
with their gas disks counter-rotating with respect to their stellar disks.
This new mechanism of star formation  suppression (or quenching)
does not depend so strongly  on
model parameters of disk galaxies, such as bulge-to-disk-ratios and gas mass fractions.
Such severe suppression of star formation is due largely to the suppression
of the gas density enhancing mechanism i.e spiral arm formation in disk galaxies with counter-rotating gas.
Our simulations also show that  molecular hydrogen and dust
can be rather slowly consumed by star formation in disk galaxies with counter-rotating
gas disks (i.e., long gas depletion timescale).
Based on these results, we suggest that spiral and S0 galaxies with counter-rotation
can have rather low star formation rate for their gas densities.
Also we suggest that a minor fraction of S0 galaxies have no prominent spiral
arms, because they have a higher fraction of counter-rotating gas.
We predict that  poststarburst `E+A' disk galaxies 
with cold gas
could have counter-rotating gas.
\end{abstract}


\vspace{-0.5cm}
\section{Introduction}

Since disk galaxies with counter-rotating components
were first discovered  in 1987 by Galletta, 
the origin of them has been discussed extensively
by observational and theoretical studies\footnote{For a list of counter-rotating galaxies refer to Corsini \& Bertola 1998 and to Corsini 2014 for a review}.
Despite the fact that counter-rotation has been observed across 
the whole Hubble sequence, counter-rotation is more frequent in early-type 
disk galaxies. 
A statistical study by Bureau \& Chung (2006)
found that (i) 60 galaxies classified as S0 have a significant amount of 
ionized gas and (ii) 14 of the S0s with ionized gas 
have counter-rotating ionized gas, 
representing (23$\pm$5)\%:
This result was 
obtained by merging samples of  Bertola et al. 1992, Kuijken et al. 1996, Kannappan \& Fabricant 2001, Pizzella et al. 2004 and Bureau \& Chung 2006. More recent study by Davis et al. (2011) have provided
strong evidence on favor of the external origin of the counter-rotating gas in early-type galaxies. While  Sarzi et al. 2006 did not rule out the internal origin of the misaligned  gas. 
Jin et al. 2016 found that about 85\% of their sample of galaxies with misaligned  gas lay in the green valley and red sequence. 
Kannappan \& Fabricant (2001) and  Pizzella et al. (2004)
have addressed the counter-rotation frequency in 
spiral galaxies with samples of 38 S0/a-Scd and 50 S0/a-Scd galaxies, respectively.
Kannappan \& Fabricant set an upper limit of 8\% on the fraction of spirals that host counter-rotating gas, whereas
Pizzella et al. (2004)
found less than 12\% of their galaxies sample hosting counter-rotating gas.

\begin{table}
\centering
\begin{minipage}{85mm}
\caption{The basic model parameters for
disk galaxies.}

\begin{tabular}{lllllll}
\hline

{Model ID \footnote{ The model M23 and 24 are the same as M1 and M2,
respectively, but star formation rate is assumed to depend on total gas density.  
}}  & 
{ $M_{\rm s}$  \footnote{ The initial total mass of a stellar disk
in units of $10^{10} M_{\odot}$.
 }} & 
{ $M_{\rm g}$  \footnote{  The initial total mass of a gas disk
in units of $10^{10} M_{\odot}$. 
}} &
{ $f_{\rm g}$  \footnote{Gas fraction, measured as the initial mass ratio of gas
to stars  in a disk galaxy.
}} &
{ $f_{\rm b}$  \footnote{  The bulge-to-disk mass-ratio in a disk galaxy.
}} &
{ $V_c$  \footnote{ `S' and `C' represents the standard rotation
(i.e., co-rotation) and counter-rotation, respectively.
}} &
{ $\rho_{\rm th}$  \footnote{The threshold gas density for 
star formation in units of cm$^{-3}$.
}} \\
M1 &  6.0 & 0.6 & 0.1  & 0.17 & S & 100 \\
M2 &  6.0 & 0.6 & 0.1  & 0.17 & C & 100   \\
M3 &  0.6 & 3.0  & 5.0  & 0 & S & 100 \\
M4 &  0.6 & 3.0  & 5.0  & 0 & C & 100  \\
M5 &  1.8 & 1.8  & 1.0  & 0 & S & 100  \\
M6 &  1.8 & 1.8  & 1.0  & 0 & C & 100  \\
M7 &  6.0 & 0.6  & 0.1  & 0 & S & 100 \\
M8 &  6.0 & 0.6  & 0.1  & 0 & C & 100   \\
M9 &  4.8 & 1.8  & 0.37  & 0.17 & S & 100   \\
M10 & 4.8 & 1.8  & 0.37  & 0.17 & C & 100 \\
M11 &  3.0 & 0.6  & 0.2  & 1.0 & S & 100  \\
M12 &  3.0 & 0.6  & 0.2  & 1.0 & C & 100  \\
M13 &  4.8 & 1.8  & 0.37  & 0 & S & 100  \\
M14 &  4.8 & 1.8  & 0.37  & 0 & C & 100  \\
M15 &  6.0 & 0.3  & 0.05  & 0.17 & S & 100 \\
M16 &  6.0 & 0.3  & 0.05  & 0.17 & C & 100  \\
M17 &  6.0 & 0.3  & 0.05  & 0.17 & S & 10  \\
M18 &  6.0 & 0.3  & 0.05  & 0.17 & C & 10  \\
M19 &  6.0 & 0.6  & 0.1  & 0.17 & S & 10  \\
M20 &  6.0 & 0.6  & 0.1  & 0.17 & C & 10  \\
M21 &  6.0 & 0.6  & 0.1  & 0.17 & S & 30   \\
M22 &  6.0 & 0.6  & 0.1  & 0.17 & C & 30  \\
M23 &  6.0 & 0.6  & 0.1  & 0.17 & S & 100   \\
M24 &  6.0 & 0.6  & 0.1  & 0.17 & C & 100  \\

\end{tabular}
\end{minipage}
\vspace{-0.15cm}
\end{table}

\vspace{-0.1cm}
Theoretical studies showed that secondary slow episodic and continuous 
infall of non-clumpy gas is the most viable mechanism 
to form massive counter-rotating disk in 
spiral galaxies according to the numerical simulation results of 
Thakar \& Ryden (1996). 
Major mergers between spirals of similar masses are unlikely
to form disk galaxies with counter-rotation, because they can completely
destroy the original disks (Thakar \& Ryden 1996; Mapelli et al. 2015;
Bassett et al. 2017).
Internal processes that may lead to counter-rotation were also discussed
by a number of authors  (e.g.,  Evans \& Collett 1994, Wozniak \& Pfenniger 1997). 
Other interesting theoretical studies found that counter-rotating stellar disks 
develop one-armed spiral waves 
due to  two-stream instability, (e.g., Lovelace et al. 1997, Comins et al. 1997).  
Furthermore,  Lovelace et al. (1997)
found that, for co-rotating stellar disk with less than 
50\% mass fraction in counter-rotating stars/gas, 
the strongest amplification occurs 
for the first mode (m = 1), which corresponds to one-armed leading waves.
Smaller amplification can be seen for the second mode (m = 2),
i.e.,  two-armed trailing waves. 

The dynamics of gas and stars in galaxies 
is a very important driver of galaxy formation
and  evolution. Accordingly, we expect that counter-rotating components,
in particular,  gaseous counter-rotation can have
significant influence on galactic stellar and gas dynamics,
 which can, in turn, influence star formation and H$_2$ formation. 
However, the above-mentioned
previous studies did not investigate this issue.
H$_2$ gas, which is the building block of molecular clouds
where star formation is ongoing,
can  be formed  on the surfaces of dust grains from neutral
hydrogen. Star formation and its feedback effects (e.g., supernova explosions)
can destroy the interstellar dust grains and therefore influence
H$_2$ formation processes.
Thus, both the formation and evolution of dust and that of molecular gas 
needs to be investigated self-consistently so that the influence of
counter-rotating gas on galaxy-wide star formation can be adequately
addressed.

Thus the purpose of this paper is to investigate whether star formation rates (SFRs) can be severely suppressed in disk galaxies, if the gaseous components are counter-rotating with respect to their stellar disks. We also discuss how the formation of molecular hydrogen (${\rm H_2}$) on dust grains can be influenced by counter-rotating kinematics of gas in detail. The physical properties
of ${\rm H_2}$ formation and star formation in counter-rotating gas disks were addressed observationally  by Bettoni et al. (2001). 
However, we present the first theoretical study on this issue in the present study. 

\vspace{-0.5cm}
\section{The model}
\begin{figure} 
\centering
\includegraphics[width=7.5cm,height=30cm,keepaspectratio]{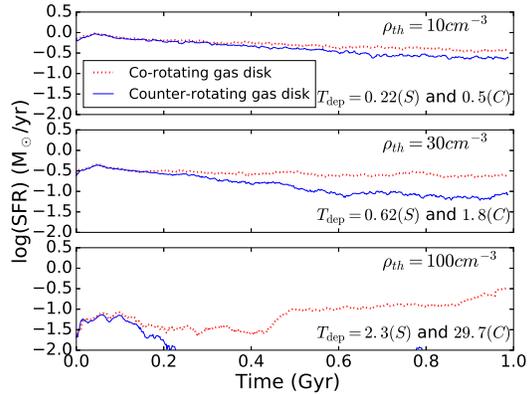}
\caption{Time evolution of SFRs in co-rotating (`S')
models (from the top) M19, M21 and M1 (red dotted) 
and counter-rotating (`C') models M20, M22 and M2 (blue solid)
for  $\rho = 10$ (top)  30 (middle), 100 cm$^{-3}$ (bottom).
Gas depletion times ($T_{\rm dep}$ ($\times 10^{10}yr$) at the initial time for co-rotating models and the final time for the counter-rotating models) 
are also indicated for each model.}
\vspace{-0.3cm}
\label{SFR_m67}
\end{figure}
In order to derive star formation rates (SFRs) of disk galaxies
with co-rotating or counter-rotating gas disks,
we adopt our original simulation code that has been recently
developed to analyze the evolution of dust, metal, molecular
hydrogen in galaxies (Bekki 2013, 2015, hereafter B13, B15).
We here describe them briefly,
because the details of the code are already given in Bekki (B13, B15).
The code is based on
the smoothed-particle hydrodynamics (SPH) method,
and it includes various physical processes of interstellar medium
and stellar evolution so that
we can investigate various processes.

\begin{figure*}
\begin{multicols}{2}
    \includegraphics[width=1.3\linewidth]{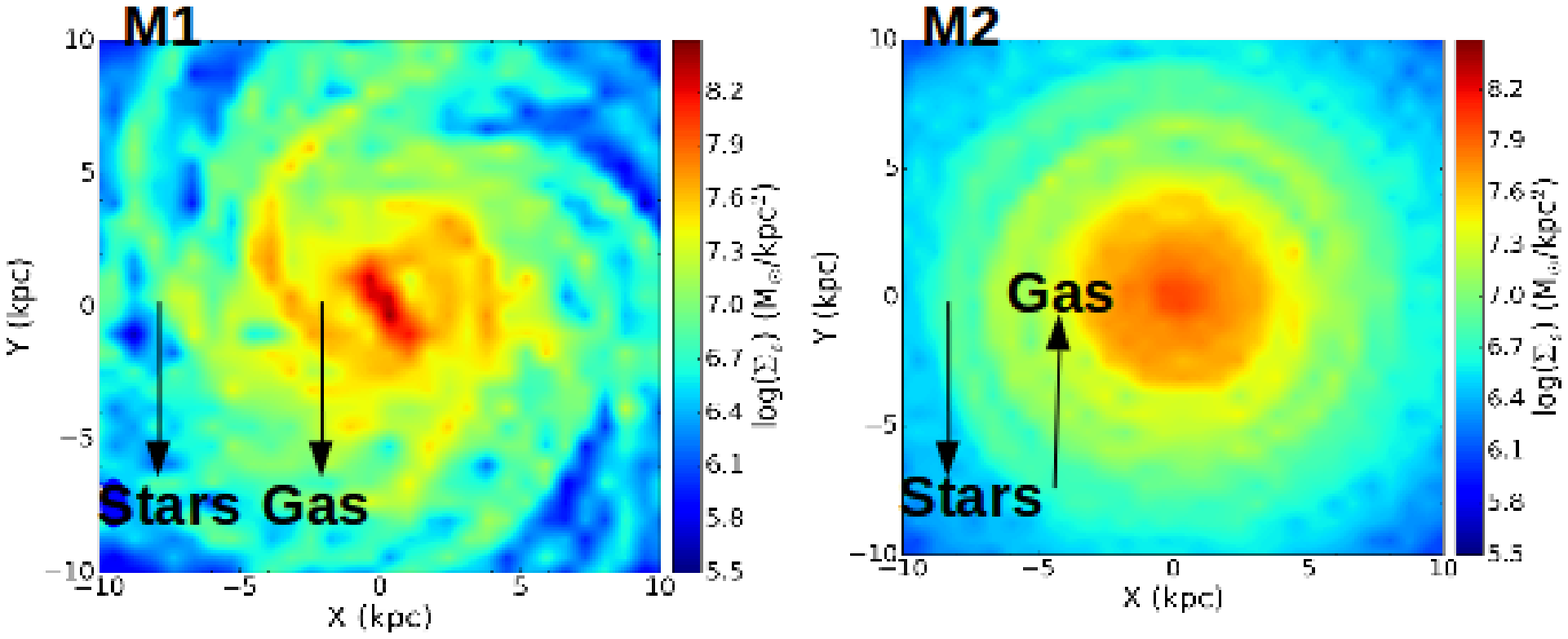}\par 
    \includegraphics[width=0.6\linewidth]{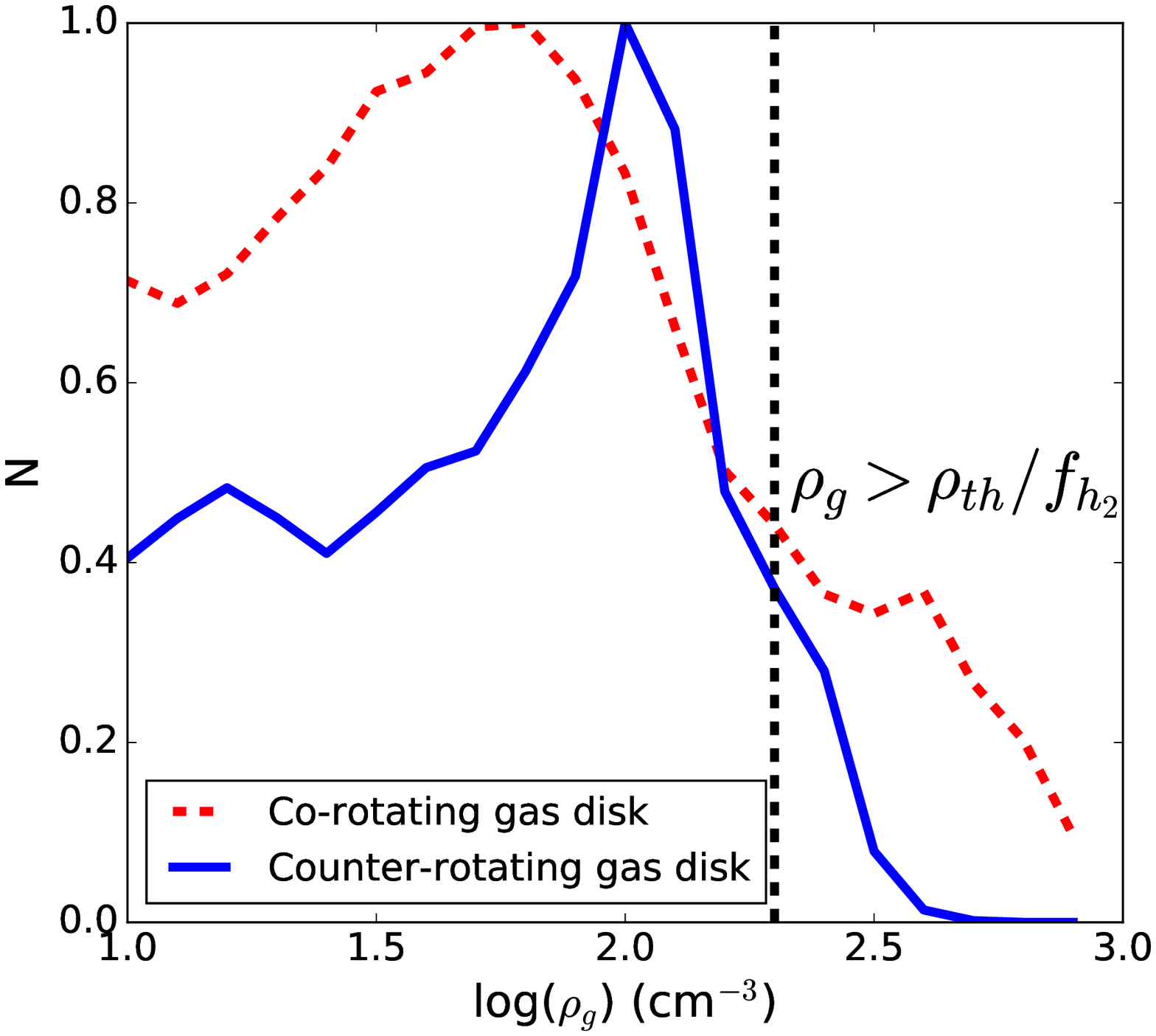}\par 
    \end{multicols}
\caption{The surface mass density of the gas disk ($\Sigma_g$, at T = 0.98 Gyr) in logarithmic scale projected onto \textit{xy} plane for the co-rotating model  M1 (left) and the counter-rotating model M2 (middle). The black arrows show the rotation direction for stars and gas. The number of particles as a function of the logarithm of the volume density (right). Red dashed and blue solid lines represent M1 and M2, respectively, and the black dashed vertical line indicates the threshold
${\rm H_2}$ gas density for star formation, in the right panel. This can be compared
with the results in Fig. 4 of  Martig et al. (2013).}
\vspace{-0.3cm}
\label{fig2}
\end{figure*}

A disk galaxy is assumed to consist of dark matter halo, stellar disk,
gas disk and stellar bulge with the total masses being
denoted as $M_{\rm h}$, $M_{\rm s}$, $M_{\rm g}$,
and $M_{\rm b}$, respectively. The
mass-ratio of gas disk to stellar disk
is denoted as $f_{\rm g}$ and considered to be a key parameter for
the time evolution of SFRs and molecular gas fraction.
The initial density profile of dark matter halo
in a disk galaxy is 
modeled using  the density distribution of the NFW
halo (Navarro et al. 1996)  suggested from CDM simulations.
The $c$-parameter ($c=r_{\rm vir}/r_{\rm s}$, where $r_{\rm vir}$
and $r_{\rm s}$ are the virial
radius of a dark matter halo and the scale length, respectively)
and $r_{\rm vir}$ are chosen appropriately
by using recent predictions from cosmological simulations
(Neto et al. 2007).

We adopt 
the Hernquist density profile for 
the bulge of a disk galaxy
and the  bulge mass fraction ($M_{\rm b}/M_{\rm s}$)
is a free parameter ranging from 0 to 1. 
The size ($R_{\rm b}$) and the mass of the bulge
has the following scaling relation that can reproduce the
properties of the Galactic bulge: 
$R_{\rm b}=2 (\frac{ M_{\rm b} }{ 10^{10} {\rm M}_{\odot} })^{0.5} $kpc.
The radial ($R$) and vertical ($Z$) density profiles of the stellar disk are
assumed to be proportional to $\exp (-R/R_{0}) $ with scale
length $R_{0} = 0.2R_{\rm s}$  and to ${\rm sech}^2 (Z/Z_{0})$ with scale
length $Z_{0} = 0.04R_{\rm s}$, respectively.
The gas disk with a size  of $R_{\rm g}$
has the  radial and vertical scale lengths
of $0.2R_{\rm g}$ and $0.02R_{\rm g}$, respectively.
Since we investigate disk models with $M_{\rm h}$ similar to
that of the Galaxy (i.e., $M_{\rm h}=10^{12} {\rm M}_{\odot}$),
the exponential disk
has $R_{\rm s}=17.5$ kpc for all models 
in the present model. The initial Toomre's parameter $Q$ is set to be  1.5 for gas and stars
and the vertical velocity dispersion at a given radius is set to be 0.5
times as large as the radial velocity dispersion at that point
so that the dynamical equilibrium in the vertical direction
can be achieved.

A gas particle is assumed to be converted into a new star, if
the local ${\rm H_2}$ density ($f_{\rm H_2} \rho_{\rm g}$
where $f_{\rm H_2}$ and $\rho_{\rm g}$ are the ${\rm H_2}$
mass function and gas density, respectively)
exceeds a threshold density
(${\rho}_{\rm th}$) and we investigate
the models with  $\rho_{\rm th} =10$, 30 and  100  cm$^{-3}$.
These are ${\rm H_2}$-dependent SF model, and we also investigate
H-dependent one in which $\rho_{\rm g}$ should be higher than $\rho_{\rm th}$.
We adopt these values, because our previous numerical simulations
of galaxy evolution with dust demonstrated that
the models with the above values can explain
the observed SFR-$\Sigma_{\rm g}$ relation
(See Fig. 3 in Yozin \& Bekki 2014 for the case of the Magellanic Clouds).
 We assume that
SFR$\propto \rho_{\rm g}^{\alpha_{\rm sf}}$ ($\alpha_{\rm sf}=1.5$)
in  the present study.
Chemical enrichment processes, dust formation and evolution in
ISM  and SN feedback effects
are included in the same way as 
B13 and 15.

We investigate disk models with co-rotating (referred to as `standard'
rotation and labelled as `S' for convenience) and counter-rotating (`C') gas disks
for a given set of disk parameters. By comparing between the two models,
we try to understand how counter-rotating gas can influence dynamics
and star formation histories of disk galaxies. 
We show the results for only 24 models in the present study and
Table 1 describes the model parameters for each of the `S'-`C' pairs.
The total number of particles used for a simulation
is 1033400 for a fiducial model (M1 and M2) and
different initial  gravitational softening lengths ($\epsilon$)
are allocated for different four components:
$\epsilon=2.1 kpc$ and 0.26 kpc for dark matter and baryonic components.

\vspace{-0.5cm}

\section{Results}
Fig. \ref{SFR_m67} shows the time evolution of the SFRs  in the co-rotating models M1, M19 and M21
and counter-rotating models M2, M20 and M22. 
All the co-rotating and counter-rotating models have the same parameters except for the SF density threshold, $\rho_{th} = 10, 30, 100  cm^{-3}$ (top to bottom). 
The SFR in all models with counter-rotating gas disk is suppressed in comparison with co-rotating gas disk models. Changing the SF density threshold only changes
the magnitude of suppression: 
As we go to higher $\rho_{\rm th}$,
SF suppression becomes stronger. This result can also be seen in models M23 and M24. Gas depletion times ($T_{\rm dep}$) are also indicated for the different models (see e.g  Martig et at 2013,  Davis et al. 2013, Davis \& Bureau 2016, for the $T_{\rm dep}$ calculation). To understand the reason behind this suppression we studied the time evolution of the gas surface density distribution ($\Sigma_g$)  at T = 0.49 Gyr and T = 0.98 Gyr.
\begin{figure} 
\includegraphics[width = 7.5 cm,height=30cm,keepaspectratio]{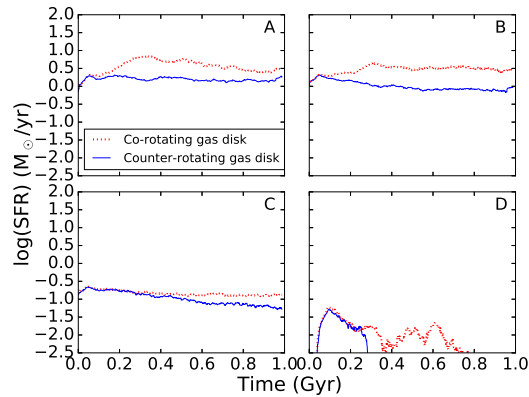}
\caption{The time evolution of SFRs in models; (A) M5 and M6, (B) M9 and M10, (C) M17 and M18, (D) M11 and M12. Red dotted lines and blue solid lines represent co-rotating and counter-rotating models, respectively.}
\vspace{-0.3cm}
\label{JointSFR}
\end{figure}
The 2D distribution of the gas disks in models M1
with  co-rotation gas (left), M2 
with counter-rotation gas (middle) and the number of particles as a function of the logarithm of the volume density of the two models (right) at time T = 0.98 Gyr are shown in Fig. \ref{fig2}. The surface mass density of the gas ($\Sigma_g$) in logarithmic scale is plotted for the two models. 
Clearly, the 2D distribution  shows the formation of gas concentrations around the 
spiral arms and within the stellar bar
for the co-rotating gas disk, whereas
 the spiral arms and bar completely fail 
to form in this case of  counter-rotating disk. Since high fraction of star formation takes place around spiral arms, due to the enhanced gas density by the arms, the lack of those arms leads to suppression in the SFRs. 
The stellar bar, which was formed as the result of the dynamical evolution of the system, can drive rapid inward transfer to the
inner region of the stellar disk so that star formation can become active there. The volume density as well shows the domination of the co-rotating model in two density regimes, below 10$^{1.8}$ and above 10$^{2.4}$ cm$^{-3}$, however, the second regime is the one relevant for star formation. The regime over which counter-rotating model is dominant, most probably, is the one in which H$_2$ is forming (see fig. 4). 

Fig. \ref{JointSFR} shows the time evolution of SFRs for co-rotating and counter-rotating gas disks in models with different model parameters. The SFR suppression for the counter-rotating gas disks is evident in all models even in the cases in which models with counter-rotating gas disks show spiral structure. 
In 60\% of the counter-rotating models in this paper,
no spiral arms or bars were formed.
The rest of the models (40\%) 
show weak spiral structures that are regarded as  `leading' arms rather
than  `trailing' ones (with respect to stellar rotation).
Models in (A) exhibit a 
kind of clumpy structure in the co-rotating case (M5) and smoother structure with spirals in the counter-rotating case (M6). 
For both cases of co-rotation M9 and counter-rotation M10 in (B),
spiral arms can be  formed, however, with smoother structure in the case of M10. Models M17 and M18 in (C) and models M11 and M12 in (D) show spiral structure for the co-rotating gas disk, while the spiral structure completely fails to form in counter-rotation case. Star formation proceeded with significantly lower rates in the model M11 with big bulge, while star formation
is truncated within $\sim 0.3$ Gyr for the model M12 with counter-rotating gas. 

\begin{figure} 
\centering
\includegraphics[width = 7.5 cm,height=30cm,keepaspectratio]{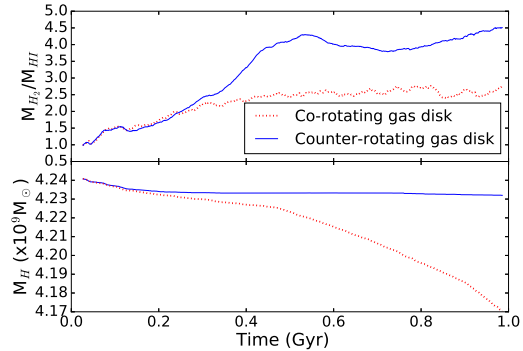}
\caption{The time evolution of the ratio of molecular hydrogen to atomic hydrogen, upper panel, and that of the total gas, lower panel. Red dotted lines and blue solid lines represent M1 and M2, respectively.}
\vspace{-0.3cm}
\label{Hydrogen1and2_m67}
\end{figure}

Previous observations revealed
physical correlations between  H$_2$ properties (e.g., mass and  H$_2$/HI ratio)
and  galactic  basic physical properties (mass, size, morphological types in
disk galaxies with co-rotating and counter-rotating 
disks  (e.g., Casoli et al. 1998). Fig. \ref{Hydrogen1and2_m67} 
shows the time evolution of H$_2$/HI ratio and the time evolution of the total gas in models M1 and M2. 
The gas consumption is slower in the counter-rotation case due to 
the lower star formation rate ( due to the lack of the mechanism that enhances gas density, i.e, the spiral arms).
Owing to the slow gas consumption in M2,  H$_2$ formation on dust grains
can continue and consequently the H$_2$-to-HI-ratio becomes higher in
M2 (M$_{H_2}$ = 3.07 $\times$ 10$^9$, 3.46 $\times$ 10$^9$ M$_{\odot}$ at T = 0.98 Gyr for models M1 and M2 respectively). 
Although possible enhancement of the HI to H$_2$ conversion 
efficiency in disk galaxies with counter-rotation gas was 
already speculated by  Bettoni et al. (2001),
the present study first demonstrates higher H$_2$-to-HI-ratios in simulated disks with
counter-rotation.
Since H$_2$ forms exclusively on the surface of dust grains
in the present model, 
its higher abundance in counter-rotating 
gas disks indicates quieter and milder environment with less 
dust destruction processes. Indeed, the final dust masses in M1 and M2
are  7.87 $\times$ 10$^7$ and 9.45 $\times$ 10$^7$ M$_{\odot}$, respectively.

\vspace{-0.1cm}

\begin{figure*} 
\includegraphics[width=13cm,height=32cm,keepaspectratio]{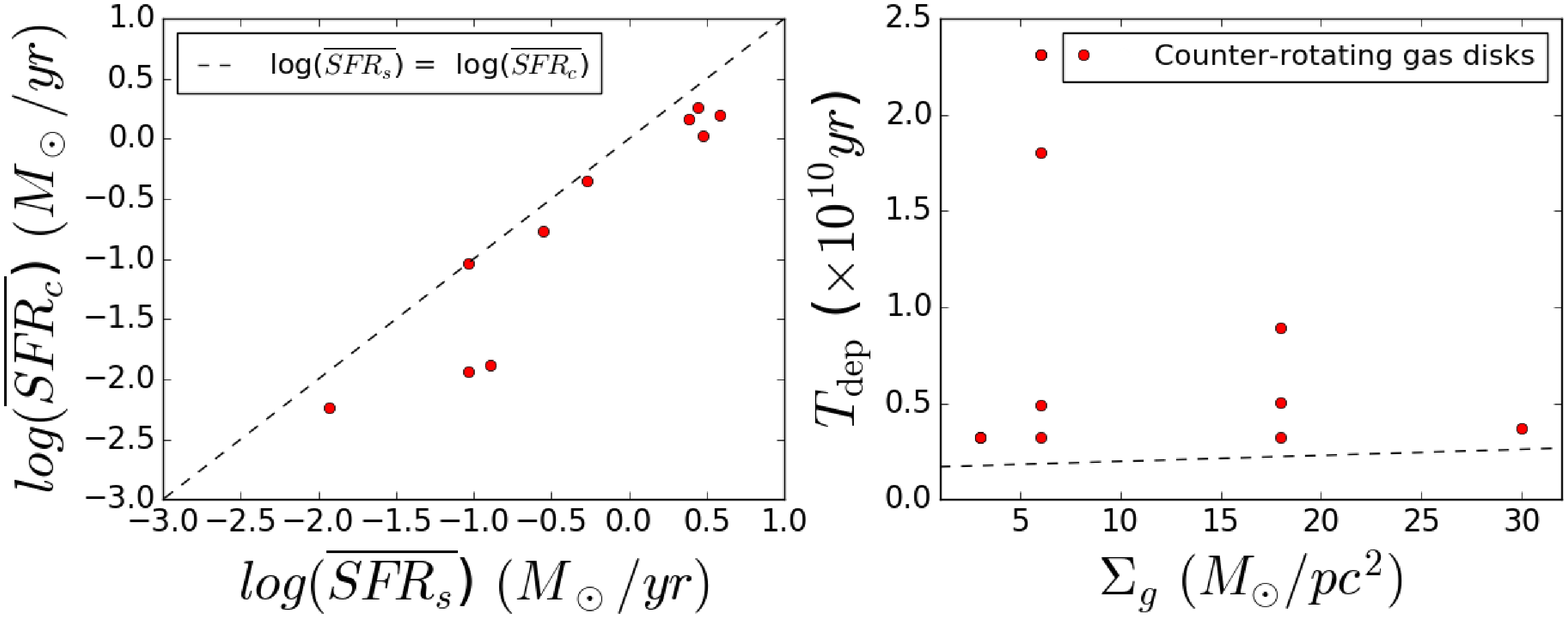}
\caption{The mean SFR ($\overline{SFR}$) for counter-rotating 
(`c') gas disk as a function of $\overline{SFR}$ of disk galaxies with 
co-rotating (`s') disk (left) and
the gas depletion time scale ($T_{\rm dep}$) as a function of $\Sigma_{\rm g}$
(right). The dashed line is $T_{\rm dep}$ = 0.17$\times\Sigma^{0.13}_{g}$, from Martig et al. 2013.}
\vspace{-0.3cm}
\label{SFRmeans}
\end{figure*}

Fig. \ref{SFRmeans} shows the mean SFR 
in disks with counter-rotating gas disks as a function of that
in disks with co-rotating gas disks. Notably, all the points lie 
below the line at which the SFR mean for the co-rotating disks 
is equal to that for counter-rotating disks. 
We have also found that these results can be seen in the models 
with big bulges ($f_{\rm b}=1$ and small gas fractions $f_{\rm g}=0.05$).
These therefore confirm that galaxy-wide star formation can be strongly
suppressed in disks with counter-rotating gas for different galaxy-types. The higher dust amount in counter-rotating gas disks, owing to the inefficient dust destruction mechanisms, allow higher amounts of HI gas to combine into H$_2$ gas. Despite this fact, the SFR is lower in all the different cases. 
This is mainly because spiral-arm formation, which can  induce
the formation of local high-density regions,  is severely suppressed in
the models with counter-rotating gas.
Fig. 5 also shows that the depletion time scale, which is 
defined as $M_{\rm H_2}/{\rm SFR}$, is longer 
in our counter-rotating models in comparison with the recent observational
results by Martig et al. (2013).

\vspace{-0.5cm}
\section{Discussion and conclusions}
There are many different mechanisms that lead to star formation quenching in galaxies,
such as  environmental effects (e.g., ram pressure stripping),
 supernovae feedback  and AGN feedback. 
Here we have shown  a new mechanism for severe suppression
of star formation in disk galaxies with counter-rotating gas disks.
To address the validity and efficiency of this mechanism,
we have investigated 24 models with different model parameters for disk
galaxies. 
We have found that suppression of star formation
can be seen in almost all models with different bulge-to-disk-ratios, gas mass 
fractions and threshold gas densities for star formation.
The derived rather  low SFRs in disk galaxies with counter-rotating
gas is due to the lack of gas density enhancing mechanism, i.e, spiral arms, 
which can trigger the formation of massive
stars in disk galaxies.
Owing to the lack of spiral arms, 
gas densities can not be locally so high in counter-rotating gas.
\vspace{-0.05cm}
In most of the present counter-rotating models ($\sim 60$\%),
both spiral arms and bars fail to form within a time scale of 1 Gyr:
these models show either very smooth density distributions or weak ring-like
structures. It is surprising that spiral arm formation is completely
suppressed in gas-poor disk galaxy models in the present study. Probably,
a physical mechanism of spiral arm formation (e.g., swing amplification mechanism)
does not operate efficiently, if the counter-rotating gas has small fraction ($\rm f_g$).
These results  strongly suggest that counter-rotation is a 
viable explanation for the lack of spiral arms in a fraction of the gas bearing S0 galaxies. 
Recent numerical simulations of S0 formation with counter-rotation
have discussed the origin of counter-rotating gas 
(e.g., Mapelli et al. 2015; Bassett et al. 2017), they did not 
discuss why spiral arms can not be generated in S0s. The present study,
for the first time,  provides a physical explanation for the lack of spiral arms in
S0s with counter-rotating gas.

\vspace{-0.1cm}
`E+A' galaxies are galaxies 
that have spectra with strong optical absorption lines but no or little optical 
emission lines, which indicates
that these E+A galaxies have a significant fraction of 
young A-type stellar population yet low star formation rate.
The importance of those systems comes from their 
suggestive transition state between disk dominated, star forming 
galaxies and quiescent, passive spheroids (e.g., Zabludoff et al. 1996).
Zwaan et al. 2013 and French et al. 2015 
looked at the gas content (HI and H$_2$, respectively) of the E+A disk galaxies. 
Zwaan et al. found that the gas fraction ranging from 0.01 to 0.1
whereas French et al. (2015) suggested
that the low SFRs in E+As 
with significant fractions of ${\rm H_2}$  can not be due to lack of gas. 
There has been some proposals to explain the presence of HI gas with the 
lack of star formation (Buyle et al. 2006, Pracy et al. 2014). 
In this study we have found that despite the existence of gas, 
star formation is suppressed severely.
Accordingly we suggest  that some E+As
with cold gas  could potentially 
have counter-rotating gas disks.

\vspace{-0.7cm}
\section{Acknowledgment}
 
We are   grateful to the referee  for  constructive and
useful comments that improved this paper.
\\

\noindent \large{\textbf{REFERENCES}}\\

\noindent Bassett R., Bekki K., Cortese L., Couch W. J., 2017, arXiv, arXiv:1704.08434\\
Bekki K., 2013, MNRAS, 432, 2298\\
Bekki K., 2015, MNRAS, 449, 1625\\
Bertola F., Buson L. M., \& Zeilinger W. W., 1992, ApJ, 401, L79\\
Bettoni D., Galletta G., Garcia-Burillo S., \& Rodriguez-Franco A., 2001, A\&A, 374, 421\\
Bureau M., \& Chung A., 2006, MNRAS, 366, 182\\
Buyle P., Michielsen D., De Rijcke S., et al., 2006, ApJ, 649, 163\\
Casoli F., Sauty S., Gerin M., et al., 1998, A\&A, 331, 451\\
Comins N. F., Lovelace R. V. E., Zeltwanger T., Shorey P. A., 1997, ApJ, 484, L33\\
Corsini E. M., 2014, ASP Conference Series Vol. 486, 51\\
Corsini E. M., \& Bertola F., 1998, J. Korean Phys. Soc., 33, 574\\
Davis T. A., Alatalo K., Sarzi M., et al., 2011, MNRAS, 417, 882\\
Davis T. A., Young L. M., Crocker A. F., et al., 2013, MNRAS, 429, 534\\
Davis T. A., \& Bureau M., 2016, MNRAS, 457, 272\\
Evans N. W., \& Collett J. L., 1994, ApJ, 420, L67\\
French K. D., Yang Y., Zabludoff A., et al., 2015, ApJ, 801, 1\\
Galletta, G., 1987, ApJ, 318, 531\\
Jin Y., Chen Y., Shi Y., 2016, MNRAS, 463, 913\\
Kannappan S. J., \& Fabricant D. G., 2001, AJ, 121, 140\\
Kennicutt, R. C. Jr., 1998, ApJ, 498, 541\\
Kuijken K., Fisher D., \& Merrifield M.R., 1996, MNRAS, 283, 543\\
Lovelace R. V. E., Jore K. P., Haynes M. P., 1997, ApJ, 475, 83L\\
Mapelli M, Rampazzo R., Marino A., 2015, arXiv, arXiv:1501.00016\\
Martig M., Crocker A. F., Bournaud F., et al., 2013, MNRAS, 432, 1914\\
Navarro J. F., Frenk C. S., White S. D. M., 1996, ApJ, 462, 563\\
Neto A. F., Gao L., Bett P., 2007, 381, 145\\ 
Pizzella A., Corsini E. M., Vega Beltran J. C., \& Bertola F., 2004, A\&A, 424, 447\\
Pracy M. B., Owers M. S., Zwaan M., et al., 2014, MNRAS, 443, 388\\
Sarzi M., Falcon-Barroso J., Davies R. L., et al., 2006, MNRAS, 366, 1151\\
Thakar A., \& Ryden B., 1996, ApJ, 461, 55\\
Wozniak H., \& Pfenniger D., 1997, A\&A, 317, 14\\
Yozin C. \& Bekki K., 2014, MNRAS, 443, 522\\
Zabludoff A. I., Zaritsky D., Lin H., et al., 1996, ApJ, 466, 104\\
Zwaan M. A., Kuntschner H., Pracy M. B., Couch W. J., 2013, MNRAS, 432, 492\\
\label{lastpage}
\end{document}